# What is superatom?

Zhigang Wang*

Key Laboratory of Material Simulation Methods & Software of Ministry of Education, College of Physics, Jilin University, Changchun 130012, China.

*e-mail: wangzg@jlu.edu.cn

**Abstract**

The term “superatom” was introduced over three decades ago to describe clusters that emulate elemental atoms. The field has long been guided by the spherical jellium model, where magic numbers arise from shell closure of delocalized electrons. This Perspective argues that delocalization, not near-sphericity, is what makes a system atom-like. It shows that superatomic shell structure persists under arbitrary point-group symmetry, that superatomicity survives as a tunable quantum state across pressurized, ionized, and chemically precompressed systems, and that the symmetry rules governing superatoms are conditional, deeper than the jellium picture admits. The future of this field lies not in finding more magic numbers, but in exploiting superatomic states as artificial quantum systems at the atomic level.

The concept was formalized in 1992[1]. It was proposed that clusters of specific size and composition could emulate the chemical properties of elemental atoms, giving cluster science a powerful metaphor. The jellium model, first applied to explain the magic numbers of alkali-metal clusters [2], guided the field for decades. It still does. But a cluster is not a sphere. It is a molecule whose valence electrons occupy molecular orbitals delocalized across the entire nuclear framework, giving rise to a shell-like electronic structure analogous to that of a single atom [3, 4]. Delocalization, not near-sphericity, is what makes a system atom-like. The jellium model is based on a uniform positive background and spherical symmetry. It is a mathematical convenience, not a physical description of real clusters. Its shell sequence is a consequence of its assumptions, not a universal rule for delocalized electrons. This Perspective argues that the jellium picture, for all its successes, has clear limits.

Over three decades, the jellium picture has been refined. It was extended to ligand-protected clusters by introducing a counting scheme for the formal valence electrons in metal clusters [5], formalized through the Jelliumatic shell model [6], adapted to non-spherical architectures through the super-valence-bond theory [7] and through $sp^3$-hybridized superatomic molecular orbitals (SAMOs) that mimic simple molecules [8], and verified experimentally via photoelectron spectroscopy [9]. A unified view of ligand-protected systems has emerged [10], and the field has since expanded into materials science [11].

Symmetry-based electronic structures of superatoms go beyond the jellium model [12]. The jellium model, derived from a uniform spherical positive background, yields a shell sequence ($1S^2|1P^6|1D^{10}|2S^2|1F^{14}|\ldots$) in which the angular quantum number $l$ is unbounded, unlike the atomic Coulomb case where $n \geq l+1$. This sequence is a mathematical consequence of the uniform-background approximation, not a universal property of delocalized electrons. When the nuclear framework breaks spherical symmetry, the true delocalized states reorganize according to point-group rules: SAMOs of the same angular momentum split into irreducible representations, and their splitting patterns are completely determined by point-group theory. Under pressure, the endohedral metallofullerene superatom $U@C_{28}$ with $T_d$ symmetry undergoes characteristic axial compressions: the triplet ground state is preserved under $D_{2d}$ symmetry compression, but switches to a singlet under $C_{2v}$ or $C_s$ compression. The system gradually evolves from a stereoscopic to a near-planar superatom, and the electron density distributions of SAMOs, such as $D_{z2}$ and $F_{z3}$, gradually contract along the restricted degrees of freedom, so that delocalization is destroyed [13]. This is not what the jellium model, built on a spherical potential, would predict for a compressed system [14, 15].

A practical criterion follows from the discussion above. A system qualifies as a superatom if its occupied valence molecular orbitals can be assigned to well-defined superatomic angular-momentum shells (S, P, D, F, and so on) via point-group symmetry analysis, regardless of whether the filling sequence matches the jellium shell ordering. This criterion is grounded in the actual electronic structure of the system, not in any approximate model.

Superatomicity survives under ionization. Nitrogen-ring anions $N_5^-$, $N_6^{4-}$, and $N_4^{2-}$ exhibit shell structures analogous to those of simple atoms [16], and ambient-stable nitrogen-ring crystals have been designed [17]. Chemical precompression also induces delocalized metal bonds in hydrogen storage crystals, such as $H_9@C_{20}$ and $H@Au_{20}$ [18, 19], and related compressed hydrogen aggregates [20]. These examples demonstrate that superatomic states can be exploited for practical function.

SAMO interactions follow conditional symmetry rules. In $H@Au_{20}$, parity-forbidden transitions become allowed under specific symmetry lowering [19], revealing that selection rules in superatoms are not fixed by an inherited jellium symmetry but are tunable properties of the electronic state.

The broader implication is that superatoms deserve to be studied not as a subclass of clusters,

but as artificial quantum systems at the atomic level, where the laws of symmetry, shell filling, and confinement can be examined and tuned. This perspective aligns with calls to move beyond the periodic table of elements by exploiting superatom-based building blocks [21]. The future of this field lies not in finding more magic numbers [22], but in exploiting superatomic states, using pressure, charge, and chemical environment to dial in specific quantum properties, with the precision that atomic physics brought to the periodic table.

## References

[1] Khanna, S. N. & Jena, P. Assembling crystals from clusters. Phys. Rev. Lett. 69, 1664–1667 (1992). DOI: 10.1103/PhysRevLett.69.1664

[2] Knight, W. D. et al. Electronic shell structure and abundances of sodium clusters. Phys. Rev. Lett. 52, 2141–2143 (1984). DOI: 10.1103/PhysRevLett.52.2141

[3] de Heer, W. A. The physics of simple metal clusters: experimental aspects and simple models. Rev. Mod. Phys. 65, 611–676 (1993). DOI: 10.1103/RevModPhys.65.611

[4] Brack, M. The physics of simple metal clusters: self-consistent jellium model and semiclassical approaches. Rev. Mod. Phys. 65, 677–732 (1993). DOI: 10.1103/RevModPhys.65.677

[5] Häkkinen, H., Walter, M. & Grönbeck, H. Divide and protect: capping gold clusters with molecular gold. J. Phys. Chem. B 110, 9927–9931 (2006). DOI: 10.1021/jp0619787

[6] Teo, B. K. & Yang, S.-Y. Jelliumatic shell model. J. Cluster Sci. 26, 1923–1941 (2015). DOI: 10.1007/s10876-015-0921-7

[7] Cheng, L. & Yang, J. Communication: New insight into electronic shells of metal clusters: Analogues of simple molecules. J. Chem. Phys. 138, 141101 (2013). DOI: 10.1063/1.4801860

[8] Muñoz-Castro, A. $sp^3$-hybridization in superatomic clusters: Analogues to simple molecules involving the $Au_6$ core. Chem. Sci. 5, 4749–4754 (2014). DOI: 10.1039/C4SC01792A

[9] Huang, W. & Wang, L.-S. Photoelectron spectroscopy of superatom clusters. Annu. Rev. Phys. Chem. 59, 235–262 (2008). DOI: 10.1146/annurev.physchem.59.032607.093646

[10] Walter, M. et al. A unified view of ligand-protected gold clusters. Proc. Natl. Acad. Sci. USA 105, 9157–9162 (2008). DOI: 10.1073/pnas.0801001105

[11] Doud, E. et al. Superatoms in materials science. Nat. Rev. Mater. 5, 371–384 (2020). DOI: 10.1038/s41578-019-0175-3

[12] Li, R. et al. Symmetry-based electronic structures of superatoms: beyond the jellium model. J. Phys. D Appl. Phys. 58, 183002 (2025). DOI: 10.1088/1361-6463/adc3f5
[13] Wang, R., Liu, Z., Yu, F., Li, J. & Wang, Z. High-pressure-induced electronic and structural transition of superatoms. J. Chem. Phys. 158, 244702 (2023). DOI: 10.1063/5.0152485
[14] Castleman, A. W. & Khanna, S. N. Clusters, superatoms, and building blocks of new materials. J. Phys. Chem. C 113, 2664–2675 (2009). DOI: 10.1021/jp806850h
[15] Reber, A. C. & Khanna, S. N. Superatoms: Electronic and Geometric Effects on Reactivity. Acc. Chem. Res. 50, 255–263 (2017). DOI: 10.1021/acs.accounts.6b00464
[16] Gong, Z. et al. High-energy nitrogen rings stabilized by superatomic properties. Adv. Energy Mater. 14, 2303446 (2024). DOI: 10.1002/aenm.202303446
[17] Gong, Z. et al. A stable superatomic nitrogen ring crystal under ambient conditions. Sci. Adv. 12, eaee5798 (2026). DOI: 10.1126/sciadv.eaee5798
[18] Liu, B. et al. Theoretical Design of Metallic Solid-State Hydrogen Storage Crystals by Chemical Precompression. CCS Chem. (2026). DOI: 10.31635/ccschem.026.202607962
[19] Wan, C. et al. Parity-forbidden superatomic molecular orbital interaction and aurophilicity induced H–Au bonding in $H@Au_{20}$. Sci. Adv. 11, eadx2053 (2025). DOI: 10.1126/sciadv.adx2053
[20] Fan, J.; Wan, C.; Liu, R.; Liu, B.; Gong, Z.; Jing, H.; Liu, S.; Wang, Z. Superatomic Hydrogen: Electron Delocalization at Reduced Pressures Compared to Metallic Hydrogen. Adv. Theory Simul. 9, e70539 (2026). DOI: 10.1002/adts.70539
[21] Jena, P. Beyond the periodic table of elements: the role of superatoms. J. Phys. Chem. Lett. 4, 1432–1442 (2013). DOI: 10.1021/jz400156t
[22] Khanna, S. N., Reber, A. C., Bista, D., Sengupta, S. & Lambert, L. The superatomic state beyond conventional magic numbers. J. Chem. Phys. 155, 120901 (2021). DOI: 10.1063/5.0062582

**Acknowledgements**

We thank Prof. Ning Wang for his helpful discussion. This work was supported by the Science and Technology Development Program of Jilin Province of China (grant no. 20250102014JC), the National Key Research and Development Program of China (grant no. 2024YFA1409900), and the National Natural Science Foundation of China (grant no. 11974136).

## Author contributions

Z.W. conceived the project, performed the analysis, and wrote the manuscript.

## Competing interests

The author declares no competing interests.

## Data Availability

This Perspective does not report new experimental or computational data. All results discussed are from the cited references, and the analysis and arguments are those of the author.

## AI assistance disclosure

During the preparation of this manuscript, the author used Tencent Yuanbao (large language model, Tencent, 2026) for discussions and for advice on text organization, format alignment to Perspective style, and language polishing. The author reviewed and edited all content, verified all scientific arguments and citations, and takes full responsibility for the publication.